\pdfoutput=1
\documentclass[%
amsfonts, amsmath, amssymb,
aps, 
prx, 
twocolumn, 
10pt, 
preprintnumbers, 
floats, 
altaffillsymbol, 
final, 
reprint, 
balancelastpage, 
flushbottom, 
noraggedfooter, 
showpacs, 
nobibnotes, 
nofootinbib, 
longbibliography, 
superscriptaddress,
]{revtex4-2}
\usepackage{Preamble}
\graphicspath{{./figures/}}

\newcommand{\maya}{\textsc{Maya }}
\newcommand{\etk}{\textsc{Einstein Toolkit }}

\newcommand{\bilby}{\texttt{Bilby}}
\newcommand{\pbilby}{\texttt{Parallel Bilby}}
\newcommand{\dynesty}{\texttt{Dynesty}}
\newcommand{\nrsur}{\texttt{NRSur7dq4}}

\newcommand{\ie}{{\it i.e.}, }
\newcommand{\eg}{{\it e.g.}, }

\newcommand{\ZZ}{{\cal Z}}
\newcommand{\LL}{{\cal L}}
\newcommand{\BB}{{\cal B}}
\def\<#1>{\mathinner{\langle#1\rangle}}

\usepackage{layouts}
\newcommand{\admM}{\ensuremath{M_0}}

\begin{abstract}

Dense environments hosting compact binary mergers can leave an imprint on the gravitational-wave emission which, in turn, can be used to identify the characteristics of the environment. To demonstrate such scenario, we consider a simple setup of binary black holes with an environment consisting of a scalar-field bubble. We use this as a proxy for more realistic environments and as an example of the simplest physics beyond the standard model. We perform Bayesian inference on the numerical relativity waveforms using state-of-the-art waveform templates for black-hole mergers. In particular, we perform parameter estimation and model selection on signals from black-hole mergers with different mass-ratio, total mass and loudness, hosted by scalar-field bubbles of varying field amplitude. We find that sub-dominant gravitational wave modes emitted during the coalescence and ringdown are key to identifying environmental effects. In particular, we find that for face-on signals dominated by the quadrupole mode, the environment is only detectable if both the ringdown and the late inspiral/early merger fall in the detector band, so that inconsistencies can be found between the inferred binary parameters and those of the final black hole. For edge-on mergers we find that the environment can be detected even if only the ringdown is in band, thanks to the information encoded in the quasi-normal mode structure of the final black-hole. 
    
\end{abstract}

\begin{document}
\title{Impact of ringdown higher-order modes on black-hole mergers in dense environments: \\ the scalar field case, detectability and parameter biases}
\author{Samson H. W. Leong}
    \email{samson.leong@link.cuhk.edu.hk}
    \affiliation{Department of Physics, The Chinese University of Hong Kong, Shatin, N.T., Hong Kong}
    
\author{Juan Calder\'on~Bustillo}
    \email{juan.calderon.bustillo@gmail.com}
    \affiliation{Instituto Galego de F\'{i}sica de Altas Enerx\'{i}as, Universidade de Santiago de Compostela, 15782 Santiago de Compostela, Galicia, Spain}
    \affiliation{Department of Physics, The Chinese University of Hong Kong, Shatin, N.T., Hong Kong}
    
\author{Miguel Gracia-Linares}
    \affiliation{Center for Gravitational Physics and Department of Physics,
The University of Texas at Austin, Austin, TX 78712}

\author{Pablo Laguna}
    \affiliation{Center for Gravitational Physics and Department of Physics,
The University of Texas at Austin, Austin, TX 78712}


\maketitle

\section{Introduction}
In the last decade, gravitational-wave (GW) astronomy has transitioned from a promising theoretical prospect to a firmly established field driven by observational discovery. With around 90 detections to date \cite{GWTC1,GWTC2,GWTC2.1,GWTC3}, GW observations have delivered invaluable leaps forward in our understanding of the Universe, ranging from the discovery of black-hole (BH) mergers with unexpectedly high masses \cite{GWTC3-pop} and the first tests of strong-field gravity~\cite{TGR_GWTC3} to the connection between neutron-star mergers and the production of heavy elements \cite{Abbott:GW170817}. This is, however, just the beginning. GW astronomy offers a unique path to unveil new physics beyond general relativity and the Standard Model (SM) of particle physics that may shed light on, for instance, the nature of dark matter. At the same time, GW signals can encode subtle signatures of the environments hosting their sources~\cite{Fedrow2017,Toubiana:AGN_LISA_2021}, that may make possible the identification of the environment properties.

Numerous studies have already explored, both theoretically and observationally, the possibility of finding signatures of modified theories of GR~\cite{TGR_GWTC3, Berti:TGR_2015, Yunes:GWTC1} and physics beyond the SM, such as ultralight bosons~\cite{proca_obs, proca_21g} and different types of Dark Matter candidates. 
In addition, while most existing studies consider vacuum-hosted GW sources, astrophysically realistic scenarios should most-likely involve the presence of environment of finite density and pressure such as gas/dust environments of varying density  e.g., active galatic nuclei~\cite{Graham:AGN_2023, Rowan:AGN_formation, Tagawa:AGN_2020, Ford:AGN_2022, Vajpeyi:AGN_2022, Grobner:AGN_rate, Barry:AGN_2018} or accretion disks~\cite{Khan:accretion, Yunes:accretion_2011, Novikov:1973_accretion}. 
For instance, some studies suggest evidence of two merging dwarf galaxies with central BHs~\cite{chandra:merger}. In all these situations, the progressive accretion of the environmental material is expected to impact the dynamics of the merging objects, which will produce effects of varying intensity in the emitted GWs~\cite{Sberna:AGN_LISA_2022, Vitor:Env_2022, Vitor:Env_2020, Vitor:Env_2014, Vitor:GW_EMRIs}.
 
The impact of dense environments in the GW emission of compact mergers is far from being un-charted territory. In fact, there exist phenomenological studies about environmental signatures on isolated BHs or EMRIs~\cite{Yunes:accretion_2011,Macedo2013:DMInspiral}, binaries in various modified gravity theories~\cite{Yunes:GW_EMRI,Berti:ST_NoHair,Healy:ST_BBH,Yagi:LISA,Cao:fR_BBH,Hirschmann:EMD} or within axion fields~\cite{Yang:axion}. There are also numerous studies about evolution of the scalar fields surrounding BHs~\cite{Wong:evolution, Alejandro:SFDM_2011} and their phenomena, such as 
superradiance~\cite{East:Superradiant,East:Superradiant2,Cardoso:KerrST,Zhang:BBH_superradiance}
and scalarisation~\cite{Cardoso:KerrST,Wong:scalarization}.
As the LIGO-Virgo-KAGRA \cite{AdvancedLIGOREF,TheVirgo:2014hva,KAGRA} collaboration (LVK) begins its fourth observation run (O4) and future detectors like Cosmic Explorer~\cite{CE_Horizon_study,US_CE,Abbott2017_CE}, Einstein Telescope~\cite{ET_science_2020,ET_design_2020,Punturo2010,Hild2011_ET} or LISA~\cite{LISA} join the GW detector network in the future, signatures of the matter-rich environments imparted on the gravitational waveform may emerge from the noisy data.

Most of the studies assessing the impact and detectability of dense environments through GWs have to date relied on phenomenological models like \eg~\cite{Vitor:Env_2020} (although see~\cite{Fedrow2017}). In this work, we present a study relying on full Bayesian inference on numerical-relativity simulations of black-hole mergers, which encapsulate all the rich phenomenology of these systems and provide the most accurate description of the emitted waves. As a first step towards more realistic simulations, we present here results based on binary BH (BBH) mergers immersed in one of the most simple types of environments one could envision and simulate, namely a scalar-field bubble, which we will denote by SBBH. Besides their technical simplicity, scalar fields are ubiquitous in modern physics also due to their rich phenomenological interests. As its existence has been suggested to be a candidate for dark matter~\cite{Magana_2012,Hui2017:ScalarDM}, the field supporting inflation~\cite{Copeland:ScalarInflation, Ratra:rollingScalar,Peebles1988:VarLambda} or accounting for dark energy~\cite{Gogberashvili2018,Bamba2012:ScalarDE,Ferreira:ScalarField,Peebles:DE,Copeland:DE_Review}. Using a state-of-the-art waveform model for BBHs occurring in a vacuum~\cite{NRSur7dq4}, we perform Bayesian parameter inference on GWs emitted by BBHs of varying mass-ratio, total mass, and inclination immersed in scalar fields of varying amplitudes. With this, we study the detectability of the environment as a function of the signal loudness and study possible parameter biases due to the omission of environmental effects.


As shown from numerous studies~\cite{Novikov:1973_accretion,Roupas:stellar_accretion,Yunes:accretion_2011}, a massive scalar field environment impacts the BBH orbit through two mechanisms. 
First, matter accretion causes a progressive increase of the total mass of the binary, altering the frequency evolution of the orbit (and therefore that of the emitted GWs) with respect to that in a vacuum. As a result, the mass of the remnant BH is larger than that corresponding to the same initial binary parameters, but in a vacuum environment. Translated to one of the most common tests of GR with GWs, this would cause an \textit{apparent violation of GR} in the so-called inspiral-merger-ringdown consistency test~\cite{TGR_IMR}, for example, see \eg~\cite{Pang:2018hjb}.
Secondly, the dynamics and dissipation of the binary will be affected by the dynamical friction exerted by the environment, which can be interpreted as the gravitational pull on the BHs by the local over-densities of the scalar field. While the previous effects can be encoded in {\it negative} order of post-Newtonian corrections, which become weaker near merger, Refs.~\cite{Valerio:tidal1, Valerio:tidal2} have shown that the specific presence of a scalar field environment can also induce non-trivial tidal effects at the {\it fifth} post-Newotnian order which become progressively stronger as the binary approaches the merger stage. The modified inspiral and merger dynamics will then affect the initial geometry of the final BH, impacting its ringdown structure. This will cause the excitation of the different ringdown modes to differ from that in a vacuum. This last aspect will become particularly useful for massive binaries, for which only the ringdown signal from the final black hole is visible in the detector band. In fact, we will show that for orientations so that the GW signal displays multiple ringdown modes, environmental effects will be detectable through the \textit{ringdown structure}, even when the inspiral signal is not observed.

The remainder of this article is structured as follows: in Sec.~\ref{sec:SBBH_NR}, we give a brief overview of the SBBH simulations we use, then in Sec.~\ref{sec:method} \&~\ref{sec:setup}, we introduce the method and the setup we use to analyze SBBH data. Finally, in Sec.~\ref{sec:result} \&~\ref{sec:discussion}, we present the results and discuss the implications and applications of such SBBH scenario.

\section{Numerical Methodology\label{sec:SBBH_NR}}

In this work, we consider a minimally-coupled massive scalar field $\phi$, with mass $m_\phi=0.4/\admM$, where $M_0$ is the initial Arnowitt-Deser-Minser (ADM) mass of the BBH. We use dimensionless quantities scaled by $\admM$. The stress-energy tensor is given by
\begin{equation}
    T_{\mu\nu} = \nabla_\mu \phi \nabla_\nu \phi - 
   \frac{1}{2} g_{\mu\nu} \qty(\nabla_\alpha \nabla^\alpha \phi + m_\phi^2 \phi^2 )\ .
   \label{eq:SE_tensor}
\end{equation}
The scalar field evolution is governed by the Klein-Gorden equation, and the space-time is evolved with the BSSN formulation~\cite{Baumgarte:NR} using the \maya code~\cite{Goodale:Cactus,Schnetter:Carpet,CarpetCode:web,Husa:Kranc,GTcatelogue}, in our local version of \etk~\cite{EinsteinToolkit:2022_11}. The time-like unit vector to the space-like hypersurfaces is evolved with the moving puncture gauge~\cite{MovingPunc:1,MovingPunc:2}.

The two BHs, modeled by punctures, are initially separated by a coordinate distance of $D = 8\,\admM$. The initial data are constructed with the York-Lichnerowicz conformal form of the Hamiltonian and momentum constraints~\cite{Lichnerowicz:YL, York:YL1, York:YL2}, which are solved using the Bowen-York solution for the momentum constraint~\cite{BowenYork, Pablo:BYNS}. The Hamiltonian constraint is solved with a version of the \textsc{2Punctures} code modified to handle scalar field environments~\cite{2Puncture}. 

The scalar field is initially vanishing at the initial location of the BHs, with its conjugate momentum $\Pi$ given by a spherical shell of radius $r_0 = 12\,\admM$ and Gaussian profile with thickness $\sigma = 1\,\admM$:
\begin{equation}
    \Pi = \Pi_0 \,\exp\qty[-\frac{1}{2} \qty(\frac{r - r_0}{\sigma})^2]~.
    \label{eq:sf_profile}
\end{equation}
We report results for different amplitudes $\Pi_0$ as $A = \Pi_0\,10^2/\admM^2$. Such a symmetric profile allows for both ingoing and outgoing scalar field modes~\cite{Alejandro:SFDM_2011}. For further details details of our simulations, we invite the reader to check Ref.~\cite{Miguel:kick}.

\begin{figure}
   \centering
    \includegraphics[width=.5\textwidth]{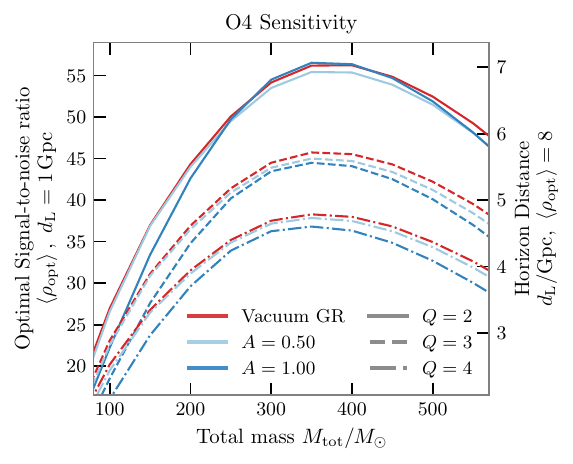}
    \caption{\textbf{Impact of scalar fields on signal loudness} We show the averaged signal loudness (denoted by optimal network SNR $\rho_{\rm opt}$) as a function of the total mass $M_{\rm tot}$ for BBHs for hosted in scalar field bubbles of varying field amplitudes in general. 
    The different curve styles represent varying mass ratios $Q = m_1 / m_2$. 
    The network SNR is computed using a detector network consisting of two advanced LIGO interferometers and Advanced Virgo working at their projected O4 sensitivity \cite{PSD_DCC, AdvLVK:prospect}, and assuming a source at a fudicial luminosity distance of \SI{1}{\Gpc}. We average the SNR over source orientations ($\theta_{JN}, \phi$) and sky location ($\alpha, \delta$) and signal polarisation $\psi$.}
    \label{fig:snr}
\end{figure}

We first assess the detectability of our target sources. Fig.~\ref{fig:snr} shows the optimal signal-to-noise ratio (SNR) (see Sec.~\ref{sec:bayes}) across our detector network of optimally oriented sources as a function of the total mass and the amplitude of the scalar field. As a general trend, we find that the presence of a scalar field reduces the SNR of the signal. This is due to the shortening of the signal caused by accelerated merger as the masses of the BHs increase when they approach each other (\eg Fig.~2 and 5 in Ref.~\cite{Yang:axion}, or Fig.~3 of Ref~\cite{Fedrow2017}). 

\section{Bayesian inference \label{sec:method}}

We perform Bayesian parameter inference on our numerically simulated waveforms for SBBHs. We inject our waveforms in zero-noise, considering an Advanced LIGO-Virgo network working at its predicted O4 sensitivity \cite{PSD_DCC,AdvLVK:prospect, AdvLIGO:sensitivity}. We do this for various scalar-field amplitudes, mass ratio, total mass, and signal loudness. The latter is characterized by so-called optimal network SNR, $\rho_{\rm opt}$ (see later). We recover our injections using the state-of-the-art model for BBHs \nrsur~\cite{NRSur7dq4}. This is a numerical-relativity surrogate model that directly interpolates through numerical simulations of BBHs including the effects of both orbital precession and higher-order harmonics. We note that our numerical SBBH waveforms contain the dominant quadruopole $(2,\pm2)$ modes plus the sub-dominant $(3,\pm 2)$, $(3,\pm 3)$ and $(4,\pm 4)$ modes while the BBH model contains all $(\ell, m)$ modes with $\ell \leq 4$. We have checked that this subtle difference in mode content does not have any significant effect on our results, both qualitatively and quantitatively.\\

\subsection{Bayesian model selection \& Parameter Estimation \label{sec:bayes}}

Assuming a waveform template model $M$, the posterior probability of the source parameters $\theta$ given a stretch of GW data $d(t)$, is given by \cite{Finn1992,Cutler1994,Romano2017}:
\begin{equation}
    p_M(\theta\,|\,d) = \frac{\pi(\theta)\,\LL_M(d\,|\,\theta)}{\ZZ_M(d)}~.
\end{equation}
Above, $\pi(\theta)$ denotes the prior probability for the parameters $\theta$ while $\LL_M(d\,|\,\theta)$ denotes the likelihood. We consider this to be the canonical likelihood for gravitational-wave transients given by 
\begin{equation}
    \LL_M(d\,|\,\theta) \propto \exp\qty[- \frac{1}{2} \<d-h_M(\theta),\,d-h_M(\theta)>]~,
    \label{eq:likelihood}
\end{equation}
where the subscript $M$ in $h_M(\theta)$ emphasises the dependence on the particular waveform model. The symbol $\<a,b>$ denotes the noise-weighted inner product~\cite{Cutler1994}, given by
\begin{equation}
    \<a,b> = 4 \Re \int_{f_{\rm min}}^{f_{\rm max}} 
    \frac{\tilde{a}(f)\,\tilde{b}^{*}(f)}{S_n(\abs{f})} \,\dd f~.
    \label{eq:nwip}
\end{equation}
Here, $\tilde{a}(f)$ denotes the Fourier transform of $a(t)$, the symbol $^{*}$ denotes complex conjugation and $S_n(\abs{f})$ is the one-sided power spectral density of the detector noise. 

Finally, the normalisation factor $\ZZ_M$ denotes the Bayesian evidence for the signal model $M$, given by 
\begin{equation}
    \ZZ_M(d) = \int \LL_M(d \,|\, \theta)\,\pi(\theta)\,\dd \theta~. 
    \label{eq:normalisation}
\end{equation}
Given two models $M$ and $N$, the relative Bayes' factor is given by $\BB_N^M = \ZZ_M / \ZZ_N$. In our particular case, the models $M$ and $N$ will correspond to the BBH and SBBH hypotheses, \ie $\ln\BB^{\rm vac}_{A} = \ln(\ZZ_{\rm vac} / \ZZ_{A})$.\\

In the large SNR limit, the likelihood is well approached by the expression $\ln\LL_M (d|\theta) \propto - \rho(\theta)^2 / 2$, where $\rho(\theta) = \frac{\<d,h_M(\theta)>}{\sqrt{\<h_M(\theta),h_M(\theta)>}}$ denotes the signal-to-noise ratio obtained when matched-filtering the data $d$ with template $h_M(\theta)$. In this study $d$ will correspond to a numerical relativity template from an SBBH $h_{M_{\rm t}}(\theta_{\rm t})$ with true source parameters $\theta_{\rm t}$. Then, the SNR can be expressed as 

\begin{equation}
    \rho(\theta) =  \rho_{\rm opt}(\theta_{\rm t})\times {\cal M}(h_{M_{\rm t}}(\theta_{\rm t}), h_M(\theta)).
\label{eq:matchSNR}
\end{equation}

Here, $\rho_{\rm opt}^2 = \<d,d> = \<h_{M_{\rm t}}(\theta_{\rm t}),h_{M_{\rm t}}(\theta_{\rm t})>$ is the optimal SNR squared, which is equal to the maximum SNR that can be extracted from the signal and therefore is used to characterized its loudness of the signal. Finally, the term ${\cal{M}}\in [0,1]$ denotes the overlap between the two waveforms, given by \cite{Cutler1994}:
\begin{equation}
    {\cal M}(d, h_M(\theta)) = \frac{\< d,h_M(\theta)>}{\sqrt{\< d,d>\< h_{M}(\theta),h_{M}(\theta)>}}
\end{equation}

\subsection{Estimating Bayesian evidences for the SBBH case through the Akaike information criterion}

Given an SBBH injection, we will compare the Bayesian evidences obtained under both the BBH and SBBH hypotheses. For BBHs in a vacuum, we have at our disposal a full waveform model that allows us to explicitly sample the likelihood over a virtually continuous parameter space. This is, however, not the case for SBBHs for which we have a handful of simulations at our disposal. Given this limitation, we can approximate the Bayesian evidence through the Akaike Information Criterion~\cite{Akaike1973:AIC}, given by
${\rm AIC}_{M} = 2k - 2\,\ln\LL^{\rm max}_{M}(d)$, where $\ln\LL^{\rm max}_{M}(d)$ denotes the maximum likelihood attained by the model $M$ and $k$ denotes its number of degrees, accounting for the Occam Penalty. The difference of AIC of different models can also serve as an analogue of Bayes' factor~\cite{NANOGrav_GWB,Juan:headon}, therefore, we can write: 
\begin{equation}
\begin{split}
 &\ln\ZZ_{A}(d_A) - \ln\ZZ_{\rm vac} (d_{\rm vac}) \simeq {\rm AIC}_A - {\rm AIC}_{\rm vac}\\ 
 &\quad=(\ln \LL^{\rm max}_{A}(d_A) - \ln \LL^{\rm max}_{\rm vac} (d_{\rm vac})) - (k_A - k_{\rm vac})\ .
\end{split}
\end{equation}

The terms $\ln \LL^{\rm max}_{A/{\rm vac}}(d_{A/{\rm vac}}) \propto - \rho_{\rm opt}^2(1 - {\cal M}) / 2$ depend on the loudness ($\rho_{\rm opt}$) of the injections $d_{A/{\rm vac}}$ and their match to the best-fitting templates. Assuming that the two injections $d_{A/{\rm vac}}$ have the same loudness, and given that, in each case, the injection is recovered with the same waveform model templates, we can consider ${\cal M} \approx 1$, which leads to
\begin{equation}
\ln\ZZ_{A}(d_A)  \simeq \ln\ZZ_{\rm vac} (d_{\rm vac})  - 1\ .
\label{eq:ZZ_approx}
\end{equation}
We note that this somewhat makes the implicit assumption that a random sampling would be able to find the true maximum likelihood point. However, this assumption is over-optimistic; since the likelihood difference between the maximum value and that recovered by the sampler depends on the aggressiveness of the sampler settings and the loudness of the injection.

We also note that, as written, the Akaike Information Criterion does omit the impact of the functional form of the Bayesian priors, which is crucial to determine the Bayesian evidence. We acknowledge that, in principle, this could have a strong impact on our results if, for instance, the true parameters for the BBH and SBBH differ significantly, so that the respective likelihoods would peak in regions of the parameter space with very different prior support. To provide an example, the head-on BBH mergers considered in Ref~\cite{Juan:headon} are about 10 times weaker than quasi-circular BBHs with the same total mass, mass ratio, and spins. Therefore, the head-on merger needs to be ten times closer than the BBH to produce the same $\rho_{\rm opt}$. Since astrophysically reasonable priors for the luminosity distance will roughly go as $d_{\rm L}^2$, it is expected that the head-on merger will be penalized by a factor of $\simeq 100$ w.r.t., the BBH case. In our case, however, Fig. \ref{fig:snr} shows that both systems have reasonably similar intrinsic loudness. 

\subsection{Analysis settings}

We sample the likelihood across parameter space through nested sampling to obtain the posterior parameter probabilities using the nested sampling algorithm \dynesty~\cite{dynesty, dynesty:code}, with 1024 live points. The injection of our numerical SBBH waveforms and the sampling process are all performed with the Bayesian Inference package \bilby~\cite{bilby1, bilby2}, in its  parallelized version, \pbilby~\cite{pbilby}.\\
 
Finally, while GW data analysis has classically been carried out through the analysis of the GW wave strain $h(t)$ data, we instead make use of the so-called Newman-Penrose scalar $\Psi_4 = \dv*[2]{h(t)}{t}$, following the approach recently described in~\cite{psi4_formal,bilby_Psi4}. This formalism avoids potential systematic errors arising during the obtention of $h(t)$ from the $\Psi_4$ SBBH waveforms outputted by our numerical simulations \cite{Pollney_Reissweig}.
In order to perform parameter inference, we transform the BBH GW strain templates, generated by the \nrsur\ model implemented in \texttt{LALSimulation}~\cite{lalsuite,swiglal}, together with the PSDs for the detector strain, into their corresponding $\Psi_4$ versions using the method described in~\cite{proca_obs,bilby_Psi4}.

\subsection{Figures of merit}

We will consider two main figures of merit. First, regarding model selection, we will quantify the detectability of the non-vacuum environment through the relative Bayesian evidence, or Bayes' Factor, between the BBH and SBBH models. We will denote this by 
\begin{equation}
\ln\beta = \ln(\frac{\ZZ_A(d_A)}{\ZZ_{\rm vac}(d_A)}) = \ln(\frac{\BB^A_{\rm N}}{\BB^{\rm vac}_{\rm N}})\ ,
\end{equation}
where $\BB^{{\rm vac} / A}_{\rm N}$ denotes the Bayes' factor of signal against the noise hypothesis, and recall that $\ZZ_A(d_A)$ is approximated by $\ZZ_{\rm vac}(d_{\rm vac}) - 1$. 
Second, we will study potential parameter biases arising from the omission of the environmental effects in the templates.\\

\section{Analysis setup\label{sec:setup}}

\subsection{Injected signals}

We inject SSBH signals of varying signal loudness $\rho_{\rm opt}$, total mass $M_{\rm tot}^{\rm inj}$, scalar field amplitude $A$, and orbital inclination $\theta_{JN}$\footnote{The orbital inclination angle $\iota\equiv \theta_{LN}$ is defined as the angle between the orbital angular momentum $L$ and the line-of-sight $N$; since, for our injections, we restrict to non-spinning BHs, this coincides with the angle $\theta_{JN}$ between the line-of-sight and the total angular momentum $J$.}. 
We control the signal loudness through the scaling the luminosity distance. The length of in-band signal, is determined by the total mass and the scalar field amplitudes. Increasing masses decrease the frequency of the waveform (e.g. that at the merger time), therefore reducing the amount of information present in the detector sensitive band. Next, increasing scalar fields accelerate the merger process. First, this leads to a shorter signal whose amplitude ramps up at a faster pace than in the vacuum case. Second, the final BH will display a final mass larger than that predicted by GR in the vacuum case and, even more importantly, a ringdown structure that will differ w.r.t. to a scenario where the same final BH forms in vacuum\footnote{As a consequence, for instance, Ref.~\cite{Miguel:kick} shows that the properties of the scalar field have a dramatic impact in the recoil velocity of the final BH.}. Finally, crucial for the latter effect, the orbital inclination controls the collection of emission modes contained in the observed signal. In particular, while face-on binaries only display the dominant quadrupole mode, highly inclined ones display a larger collection of modes that provides richer information on the geometry and dynamics of the source \cite{CalderonBustillo:2018zuq,CalderonBustillo2020} and, therefore, will help to distinguish between the SBBH and BBH cases.\\
 
We consider non-spinning BH mergers surrounded by scalar fields with amplitudes $A^{\rm inj} \in \{0.0, 0.5, 1.0\}$ and mass ratios $Q^{\rm inj}=m_1 / m_2 \in \{2,3,4\}$. We note that the largest scalar-field energies ($A = 1.0$) attained amongst all these cases are always roughly 11\% of the total ADM energy of the initial system~\cite{Miguel:kick}. Our injections span a grid in total (red-shifted) mass $M_{\rm tot}^{\rm inj} \in[200, 500]\,M_\odot$ separated by steps of $50\,M_\odot$.\\  

Finally, we consider 5 different signal loudness. These range from $\rho=8$, corresponding to the weakest GW detections, to $\rho=100$, which is consistent with expectations for third-generation detectors. In all cases, we performed injections for both face-on and edge-on orientation ($\theta_{JN} = 0,\;\pi/2$ respectively). Altogether, these amount to a total of 756 injections.

\subsection{Priors and sampling}

For the vanilla vacuum BBH template, we place uniform priors on the mass ratio $q = m_2/m_1 \in (1/6, 1)$, total (redshifted) mass $M_{\rm tot} \in (150, 550)\,M_\odot$, spin magnitudes $a_{1,2} \in [0,1]$, and geocentric time $t_c$. We place isotropic priors on the spin angles $(\theta_{1,2},\ \phi_{12})$, source orientation $(\theta_{JN},\ \phi,\ \psi)$, and sky localisation $(\alpha,\ \delta)$. The prior of luminosity distance is set to be uniform in co-moving volume, assuming the flat $\Lambda$CDM cosmology with Hubble parameter from Planck 2015 data --- \SI{67.74}{\kilo\meter\per\second\per\Mpc}~\cite{[][{ (See Table IV, column ``TT, TE, EE + lowP + lensing + ext'')}] Planck:2015}. We note that the mass-ratio range is determined by the domain of validity of the \nrsur{}~\cite{NRSur7dq4} waveform model and the minimum total mass is set to avoid exceeding its waveform length of $4200\,M_{\rm tot}$ of \nrsur{} at the chosen minimum frequency for the analysis, which we set to 11\,Hz. Finally, for the nested sampling, we used 1024 live points and set max. MCMC \num{10000} steps.  

\section{Results\label{sec:result}}

We now move to the description of our main results. This section is divided into two sub-sections which respectively focus on face-on and edge-on binaries. For each of them will first start assess the detectability of the scalar field as a function of the source parameters to then describe parameter biases.\\

\subsection{Face-on binaries \label{sec:face-on}}

\begin{figure*}[ht!]
    \includegraphics[width=\textwidth]{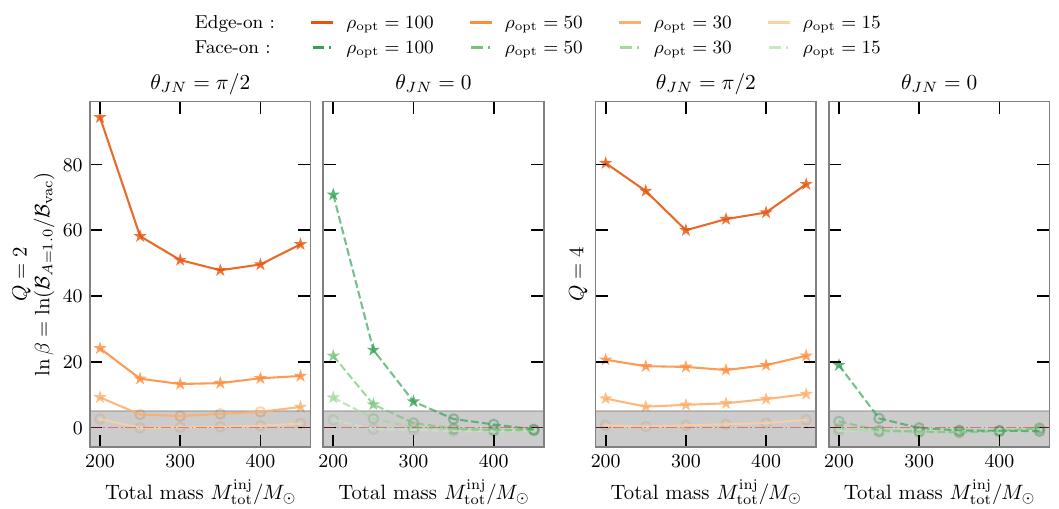}
    \caption{\textbf{Detectability of scalar fields around black-hole mergers.} The panels show $\ln\beta=\ln(\BB^{A}_{\rm N}/\BB^{\rm vac}_{\rm N})$ for edge-on (orange, solid) and face-on (green, dashed) SBBHs with $A=1.0$, plotted as a function of the total mass $M_{\rm tot}$. The grey area denotes $\ln\beta \leq 5$ region, below the detection threshold. The loudness of the injection is represented by the intensities of the color. Face-on SBBH binaries with $M_{\rm tot} \geqslant 250\,M_\odot $, are mostly indistinguishable from vacuum binaries even at $\rho_{\rm opt} = 50$.
    The two left (right) panels correspond to binaries with mass ratio $Q = 2\ (Q = 4)$. A larger $Q$ leads to less a decreased detectability due to the shorter signal. On the contrary, for edge-on orientations, a larger $Q$ favors the environment detectability due to the stronger higher harmonics. The stars (circles) on each point indicate $\ln\beta \geqslant (<) 5$. 
    } 
    \label{fig:A1_FO_vs_EO}
\end{figure*}

\subsubsection{Scalar-field detectability}

The panels labelled as $\theta_{JN}=0$ in Fig.~\ref{fig:A1_FO_vs_EO} show $\ln\beta$ as a function of the total mass, for a scalar field of magnitude $A = 1.0$.  The different curves denote different signal loudness. The grey area corresponds to $\ln \beta < 5$, which we use as the threshold for the detection of the scalar field ~\cite{Kass:BF_levels}. Finally, the two sets of panels correspond to cases with mass ratios $Q=2$ and $Q=4$.\\  

The first obvious feature for face-on binaries is that $\ln\beta$ decreases for increasing total mass. For low masses, the presence of both (a little) inspiral and merger signal in the detector band makes it possible for the vacuum BBH model to ``detect'' inconsistencies between the initial and final masses due to the accretion of scalar field. However, as the total mass increases, the inspiral (and merger) signal disappears progressively from the detector band, leaving only a single (and simple) black-hole ringdown mode, which can be perfectly captured by the vacuum model at the cost of biasing the initial parameters (mostly the total mass) of the binary. For all masses, the scalar field is tougher to detect for the $Q=4$ case due to the shorter inspiral signal. As an example, for ${\rm SNR}=100$, the field is still detectable for $M=300\,M_\odot$ for the $Q=2$ case while this is not detectable for $M=250\,M_\odot$ for $Q=4$.

\begin{figure*}[ht]
    \centering
    \includegraphics[width=\textwidth]{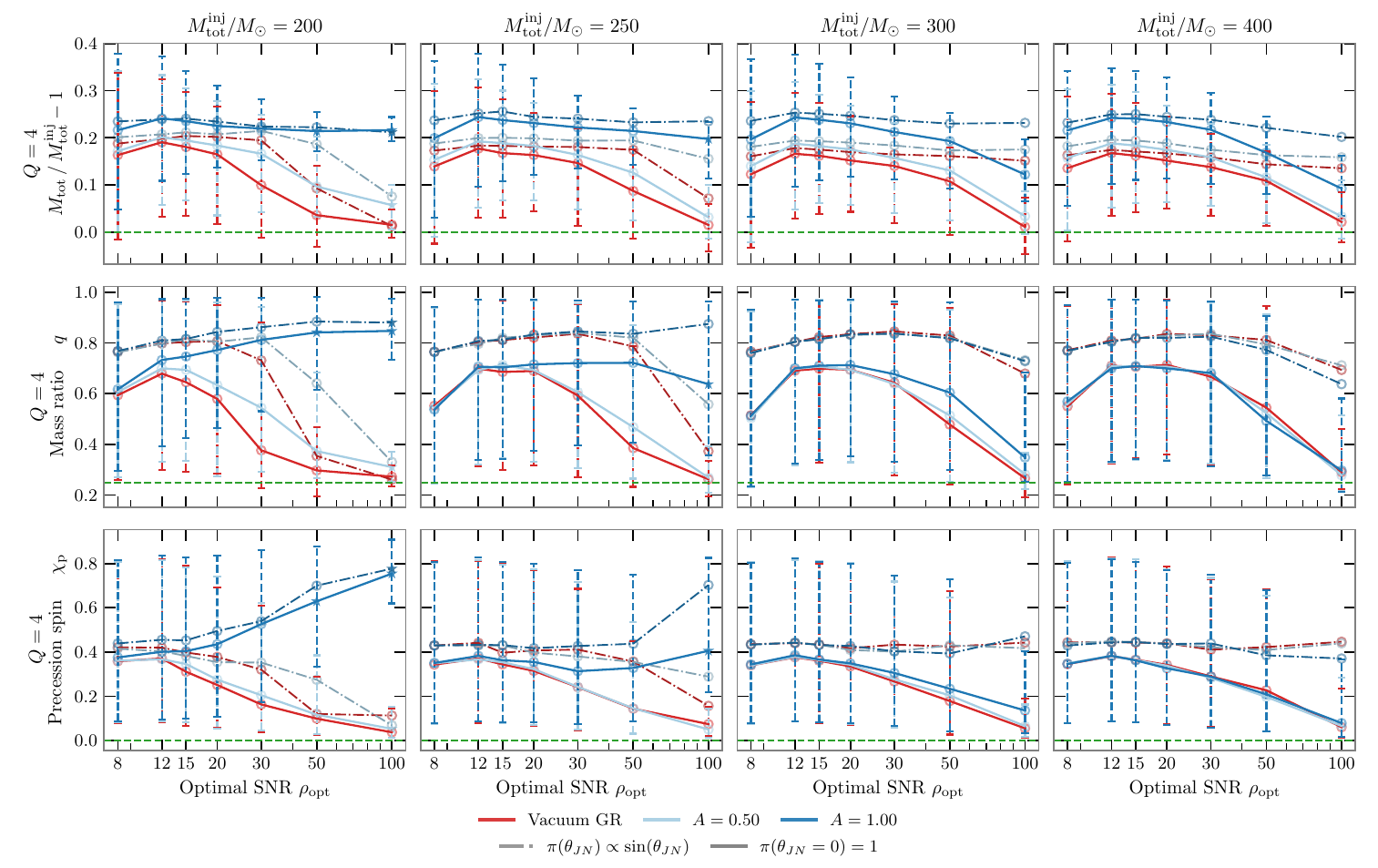}
    \caption{\textbf{Parameter biases for face-on binaries due to the omission of environmental effects.} We show fractional difference between the injected and true total mass (top panels), together with the recovered mass ratio (middle) and effective-spin precession parameter $\chi_{\rm p}$ (bottom) for our $Q = 4$ face-on injections, as a function of the signal loudness. Different columns correspond to different injected total masses, namely $M_{\rm tot}^{\rm inj} / M_\odot= \{200, 250, 300, 400\}$. The solid curves correspond to the median values of posteriors recovered with $\theta_{JN}$ fixed to 0, and the errorbars (in dashed) denote the corresponding 90\% credible intervals. The dashed-dot curves denote the median values recovered using the usual isotropic orientation prior (the errorbars are not shown for this case). The three colours (red, light blue and blue) respectively correspond to field amplitudes  $A = \{0.0, 0.5, 1.0\}$. The horizontal dashed green lines denote the true injection values. Points that are marked with a star implies $\ln\beta > 5$ for that configuration.} 
    \label{fig:faceon_bias}
\end{figure*}

The dashed curves in Fig.~\ref{fig:faceon_bias}, show the aforementioned bias in the total mass as a function of the SNR, together with those for the mass ratio $Q$ and the precession spin $\chi_{\rm p}$\footnote{The effective precession spin parameter $\chi_{\rm p}$ is defined as $\chi_{\rm p} = \max\qty(a_1\,\sin\theta_1,\frac{4q+3}{4/q+3}a_2\,\sin\theta_2)$, where $q = m_2 / m_1$ and $a_i,\theta_i$ are the spin magnitudes and tilt angles of the component BHs~\cite{chip_definition}.}, for the face-on $Q=4$ case mentioned above. Different panels correspond to different total masses. Within each panel, different colors denote various values of the scalar-field amplitude $A$, with red denoting vacuum ($A=0$). The bars delimit the symmetric 90\% credible interval (C.I.) around the median.\\

\subsubsection{Parameter biases in the absence of a scalar field}

Before discussing in detail the biases arising from the omission of the scalar field in the recovery templates, and in order to isolate its impact, we first focus on the vacuum cases. We note that unlike for the case of long signals from low-mass systems, the small (or even null) number of inspiral cycles present in the high-mass signals we analyse here can make parameter inference be dominated by Bayesian priors, leading to biased results.\\ 

To showcase the above, we note that the red curves in Fig. \ref{fig:faceon_bias} show the true source parameters are almost always outside the $90\%$ credible intervals for SNRs below $\simeq 20$, with this value increasing for increasing total mass. The reason is that within the mass range considered, very little inspiral signal falls within the detector sensitive band, making the ``observable signal'' be dominated by the emission of the final BH. In this situation, we only retrieve information about the final mass and spin. In contrast, the inference of the parent BHs is highly dominated by the Bayesian priors. In more physical terms, given the information about the final mass and spin, GR determines all the possible parent binaries, which are further weighted by our prior choices. Since our prior on the mass ratio (equal to the one regularly used by the LVK) strongly prefers equal-mass binaries, this causes a systematic ``bias'' towards such mass ratios. Next, since equal-mass binaries are intrinsically louder than unequal-mass ones, these require a larger initial mass to lead to the same final BH, which leads to a bias of the total mass towards larger values. Finally, the $\chi_{\rm p}$ parameter peaks at $\simeq 0.4$, where the prior peaks.\\ 

The above biases disappear for sufficiently high SNR, as the details of the waveform become more observable, but only for total masses below $300\,M_\odot$, when sufficient inspiral-merger information is still in band. This information be contained in either the actual inspiral signal or, in principle, in sufficiently loud overtones of the fundamental ringdown mode \cite{Giesler:2019uxc}, whose relative amplitudes and phases encode the properties of the parent binary \cite{Bustillo2021,JimnezForteza2020_overtones}. For masses above $300\,M_\odot$, the signal is completely dominated by its ringdown part making it essentially impossible to retrieve information about the binary parameters. We find that this is strongly influenced by the inability of the waveform model to resolve the inclination of the source (which, in the absence of higher-order modes is degenerate with the luminosity distance \cite{Graff:Missing_Link,London:2017bcn,CaldernBustillo2021_H0}). In fact, we find that if we restrict the inclination parameter to its true value, posterior distributions for the total mass and mass ratio progressively converge to the injected values for increasing SNR in all cases (red solid curves). While we cannot explicitly check this, we conjecture that this is due the information encoded in the overtones, as mentioned above. 

\subsubsection{Parameter biases in the presence of a scalar field}

The presence of a scalar field has two obvious consequences. First, it leads to the aforementioned bias to larger masses due to the accretion of mass by the BHs. Second, the acceleration of the merger process causes a fast ramp up of the signal amplitude close to merger that cannot  be reproduced by aligned spin systems. Instead, this ramp up can be mimicked by the modulation of the signal caused by orbital precession~\cite{Juan:headon}, which leads to a bias to large values of $\chi_{\rm p}$ for low enough masses, for which some inspiral signal is in band. We note this is analogous to the degeneracy between orbital eccentricity and precession for high-mass binaries described in~\cite{Juan:headon} which, for instance, causes this double interpretation for the GW190521 signal \cite{GW190521D,GW190521I,Gayathri2022_ecc_natastro,Gamba2022_ecc_natastro,proca_21g}. For larger masses, for which no inspiral is visible, we recover the situation for $A=0$ described before, so that $\chi_{\rm p}$ simply tends to follow its prior distribution.\\

Finally, for sufficiently low mass binaries, we find that the shortening of the inspiral signal due to the scalar field translates into an expected bias towards lower mass ratios which, consistently, increases as the SNR increases. Interstingly, while this trend is clear for the stronger $A=1$ field, such bias tends to disappear for the weaker $A=0.5$ one. For large total masses, when the shortening of the inspiral singal is no longer visible, we retrieve again the systematic bias due to the Bayesian priors described for the vacuum BBH case. 

\subsection{Edge-on binaries}

We now move to the description of our results for edge-on mergers. The signals from these systems differ qualitatively from those of face-on systems due to the presence of multiple GW modes. This has two main consequences. 

First, the presence of multiple modes in the ringdown and, in particular, their relative amplitudes and phases, depend on the initial geometry of the distorted BH formed at merger, which is determined by the properties of the parent binaries \cite{Bustillo2021,Max_GW190521,Kamaretsos2012_hms_ringdown,Hughes2019_hms_ringdown,Li2022_hms_ringdown}. This information will overcome the biases due to Bayesian priors described in the previous section. On the one hand, this can be exploited to identify the presence of a scalar-field environment at merger. On the other hand, this allows to retrieve information about the properties of the binary even at the late stages of the signal. 

Second, higher-order modes break the well-known degeneracy between the luminosity distance and the inclination of the source~\cite{Graff:Missing_Link,CaldernBustillo2021_H0}. 

\begin{figure}[hb]
    \centering
    \includegraphics[width=0.48\textwidth]{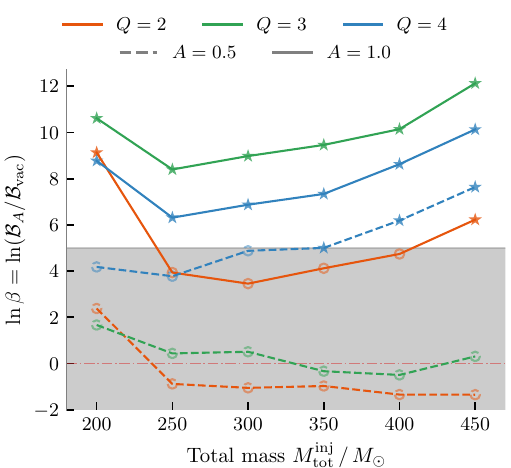}
    \caption{{\bf Scalar field detectability for edge-on SBBHs with $\boldsymbol{{\rm SNR}=30}$} Similar to the right panels (with orange curves) of Fig.~\ref{fig:A1_FO_vs_EO}, but for $\rho_{\rm opt} = 30$ with varying mass ratio and two field amplitudes ($A \in \{0.5, 1.0\}$).}
    \label{fig:EO_snr30_masses}
\end{figure}

\subsubsection{Scalar-field detectability}

The panels labelled as $\theta_{JN}=\pi/2$ in Fig.~\ref{fig:A1_FO_vs_EO} show $\log \beta$ for edge-on binaries with $Q=2$ and $Q=4$ as a function of the total mass and the SNR of the signal. These clearly show a trend as a function of the total mass that dramatically differs from the corresponding $\theta_{JN}=0$ panels. While for the latter, the field is undetectable for high mass, the detectability of the field increases for increasing mass in the edge-on cases. As said before, this is due to the increasing impact of the several higher harmonics present in the signal, which allow to detect that the ringdown structure (\ie the relative amplitude and phase of the different modes) is not consistent with those of a BH formed from a circular BBH in vacuum. This effect is extremely powerful: while for face-on $Q=2$ ($Q=4$) binaries, the scalar field is not detectable for masses above $350\ (200)\,M_\odot$ even for SNRs of 100, the field is detectable for all masses for SNRs above 50 (30) for edge-on binares. Moreover, while for face-on binaries the detectability of the field decreases for increasing mass ratio, the converse happens for edge-on ones.\\

To complement the above results, Fig.~\ref{fig:EO_snr30_masses} shows $\log \beta$ for cases with $Q=2,3,4$ and field amplitudes of $A=0.5,1.0$ at the fixed SNR of 30. First, we note that for $A=0.5$ the detectability of the field increases with increasing mass ratio. However, for the stronger $A=1$ field, we observe that the field is more detectable for $Q=3$ than for $Q=4$. For the lowest masses, we attribute to the excessive shortness of the $Q=4$ inspiral, which leads to a lower ability to identify the inconsistency between the inspiral and the final parameters of the system. Also, we understand this shorter inspiral stage, provides less time for accretion of scalar field before merger. This makes the corresponding dynamics (and therefore the initial geometry of the final BH) differ less from the vacuum case than in the $Q=3$ system, therefore leading to a ringdown structure that differs less from the vacuum case.\\

Finally, for the lowest $Q=2$ cases, the discrepancy between the final and initial parameters due to the accretion of the field is more effective at identifying the scalar field case from the vacuum case than the quasi-normal ringdown structure. The situation is reverted for the higher-mass ratio binaries, for which the reduced inspiral length combined with the richer ringdown structure, make the latter to be a better smoking gun for the non-vacuum environment.\\

The above results display, once again, the scientific potential residing in the detection of higher-order modes in GW signals from compact binaries. In addition, as we show later (and consistently with results shown by Graff~\cite{Graff:Missing_Link}), the presence of multiple modes will help to prevent parameter biases due to Bayesian priors for high-mass mergers.

\begin{figure*}
    \centering
    \includegraphics[width=\textwidth]{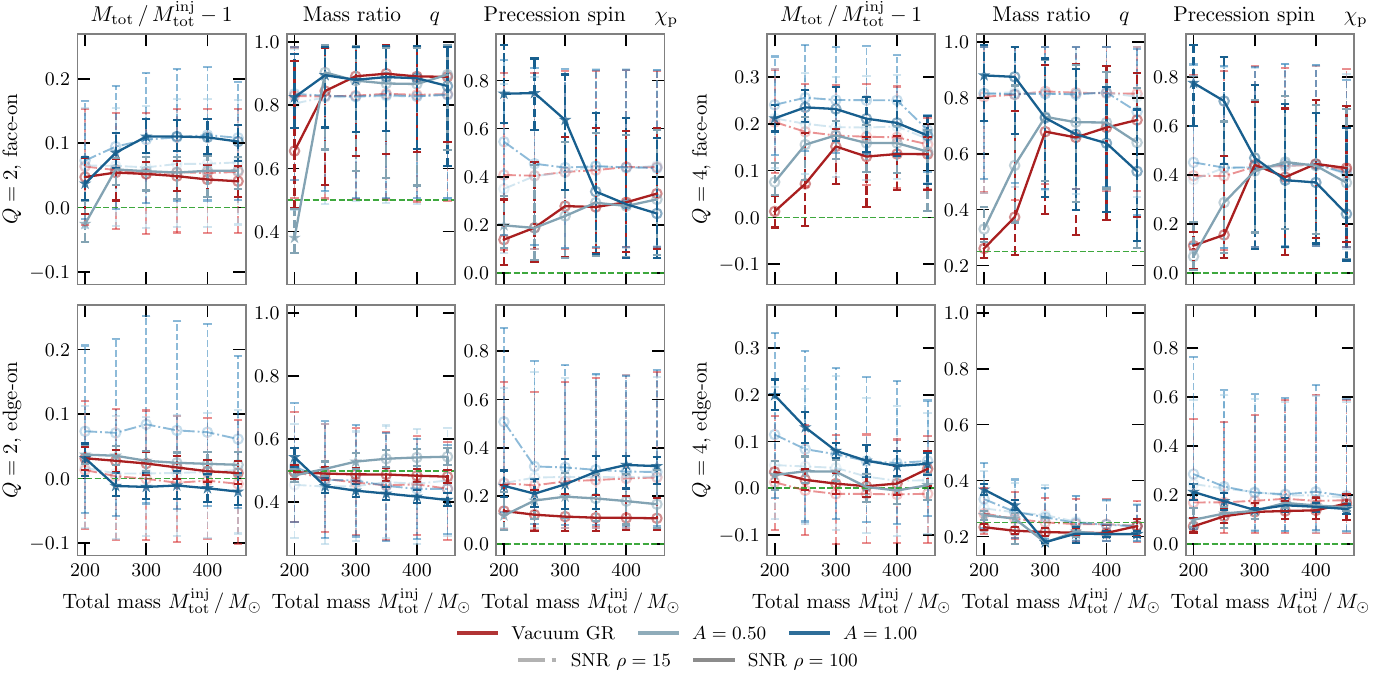}
    \caption{\textbf{Comparing parameter biases from edge-on and face-on binaries.}  
    We show the parameter biases shown in Fig. \ref{fig:faceon_bias} for the case of $Q=2$ (left) and $Q=4$ (right) SSBHs, oriented face-on (top) and edge-on (bottom) as a function of the total mass. We show these for optimal SNRs of 15 (dashed) and 100 (solid) and scalar field amplitudes (see color code). The green dashed line denotes the true injected value. 
    }
    \label{fig:comparison_fo_eo}
\end{figure*}

\subsubsection{Parameter biases}

Fig.~\ref{fig:comparison_fo_eo} shows our parameter recovery, as a function of the total mass, the field amplitude, and the signal SNR, for our $Q=2$ (left panels) and $Q=4$ cases (right panels). The bottom row corresponds to the recently discussed edge-on binaries, while the top ones correspond to face-on cases discussed previously.\\ 

First, for the vacuum case ($A=0$), we note that the main difference between face-on and edge-on cases, is that the latter return values that are way closer to the injected ones, due to the information encoded in the higher-order modes present in edge-on signals. As expected, this effect is more pronounced for the $Q=4$ case due to the stronger impact of these modes for increasing mass ratio \cite{Bustillo:2015qty,Bustillo:2016gid,Varma:2014jxa,Varma:2016dnf,Capano:2013raa,Chandra2020_Nuria}. To give an example of the magnitude of impact of the higher modes, for the $Q=4$ face-on case, we obtain a mass ratio of $\simeq 1.7$ for the highest masses, even if we set the SNR to 100. In contrast we recover the true mass ratio for the edge-on case even for an SNR of 15. The correct estimation of the mass ratio avoids the over estimation of the intrinsic luminosity discussed for face-on cases. This leads to the obtention of non-biased estimates of the total mass. Finally, while for the highest masses $\chi_{\rm p}$ is still impacted by Bayesian priors due to its almost null impact in the ringdown stage of the waveform, we note that we obtain values much closer to the true value $\chi_{\rm p}=0$ than in the face-on cases.\\

Having discussed the ``benchmark'' case of $A=0$, we move now to the discussion of the impact of the scalar field.  We note that unlike in the face-on cases, for edge-on ones it is highly non-trivial to predict the impact of the field in parameter recovery. In principle, all individual modes will suffer the same modifications as the $(2,2)$ mode in terms of their frequency and amplitude evolution. However, these will be activated with different amplitudes and relative phases than in the vacuum case during the merger and ringdown phases. This makes the resulting net signal display a complex morphology that is difficult to qualitatively relate to the vacuum scenario. Moreover, the signal morphology will strongly depend on the azimuthal angle of the observer around the source~\cite{CalderonBustillo:2018zuq,Bustillo:GW190412_kick}, which varies across our injection set. In particular, Ref.~\cite{Bustillo:GW190412_kick} showed that the signal morphology depends strongly on the angle subtended by the line-of-sight and the direction final recoil direction of the final black hole, which is different for the $Q=2$ and $Q=4$ cases. Therefore, we understand that our results should not be taken as representative of generic cases. We understand this explains (at least partially) why we observe no biases in the total mass for the $Q=2$ while $Q=4$ cases show small biases towards larger masses. In any case, we note that any bias we observe for edge-on cases is significantly smaller than those for edge-on ones. In particular, once again, the mass ratio is perfectly retrieved and $\chi_{\rm p}$ is way closer to $\chi_{\rm p}=0$.

\subsection{Larger initial separation, $D = 9\,\admM$}

\begin{figure}
    \centering
    \includegraphics[width=.48\textwidth]{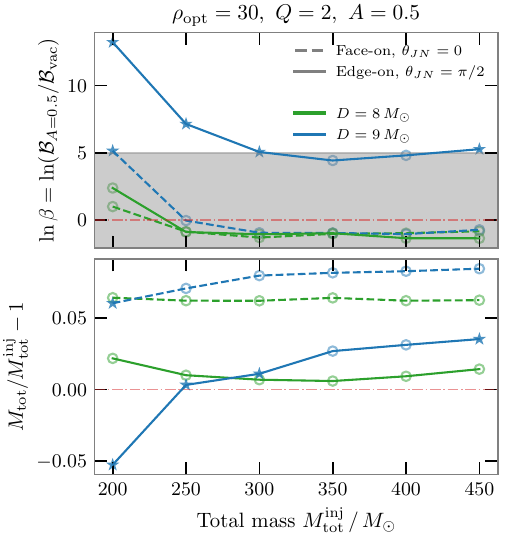}
    \caption{ \textbf{Impact of initial separation} The top panel shows in blue the $\ln\beta = \ln(\BB_{\rm A} / \BB_{\rm vac})$ as a function of injected total mass, for a case with $A=0.5$ and initial separations $D = 9\,\admM$ and $D = 8\,\admM$. The solid and dashed curves represent edge-on and face-on cases respectively. In the lower panel, we show instead the median value of the fractional bias in the total mass $M_{\rm tot} / M_{\rm tot}^{\rm inj} - 1$. Configurations which (do not) cross the threshold $\ln\beta > 5$ are denoted by a circle (star). 
}
    \label{fig:D8_vs_D9}
\end{figure}

As described in our introductory sections, the results so far correspond to cases where we start our simulations at a BH separation radius of $D = 8\,\admM$. Here, we test the impact of slightly increasing this to $D = 9\,\admM$. Our goal is to check that we obtain results consistent with the intuition built from previous sections. In principle, the longer inspiral duration should lead to a larger accretion of scalar field during a longer time, making the environment easier to detect. We repeat the same analysis as before but using a set of simulations with a separation distance of $D = 9\,\admM$ and scalar field amplitudes ranging $A \in [0.0, 0.7]$ in steps of 0.1, fixing the mass ratio to $Q = 2$.\\

The top panel of Fig.~\ref{fig:D8_vs_D9} shows $\ln\beta$ for the cases of $D = 8\,\admM$ and $D = 9\,\admM$ and $A = 0.5$, for a signal loudness of $\rho_{\rm opt} = 30$. We show results for both face-on and edge-on cases. As expected, we find that a larger initial separation greatly facilitates the detectability of the field. For instance, while for face-on cases with $D = 8\,\admM$, the field is not detectable for any total mass, this is detectable for the lowest mass considered in the $D = 9\,\admM$ case. Similarly, in the edge-on case, with a shorter separation, the field is not detectable for any total mass, but most of them are detectable or close to the threshold throughout the whole mass range in the $D = 9\,\admM$ case with $\log\beta$ comparable to the of the $A=1$ cases with separation of $D = 8\,\admM$.\\

Finally, in the bottom panel, we show explicitly that the mass accretion during the longer separation would in general lead to a larger mass biases than the $D = 8\,\admM$ counterparts, by a few percents in both face-on and edge-on cases.

\section{Conclusions\label{sec:discussion}}

Gravitational waves from compact object mergers carry information about the properties of their environments. While for current detector sensitivities, such imprints are expected to be non-detectable, the future increase in sensitivity of our gravitational-wave networks shall eventually grant access to the properties of dense environments. In this study, we have considered black-hole mergers hosted in the simplest type of dense environment one can devise, a scalar field bubble surrounding initially a BBH. First, we  performed full numerical relativity simulations of these configurations. Second, we have assessed the detectability of the dense environment in the emitted waves through Bayesian inference using the state-of-the-art waveform model for black-hole mergers \texttt{NRSur7dq4}. Finally, we have assessed parameter biases arising from the omission of environmental effects.\\

We have shown that environmental effects are detectable in two main ``qualitative ways''. First, through the inconsistency between the initial binary parameters and the properties of the final black hole predicted in a vacuum environment; provided that both inspiral and merger-ringdown are observable in the detector band. Second, through the ringdown structure of the final black hole, which differs from that of a black hole formed in a vacuum merger; provided that the inclination of the source is such that more than one ringdown mode is in band. Finally, we have shown that for face-on orientations (consistent with most detections to date) the omission of the dense environment in waveform templates roughly leads to biases towards larger total masses, equal-mass ratios and high values of the effective-spin precession parameter $\chi_{\rm p}$. This is similar to the well-known degeneracy between orbital eccentricity and precession for high-mass compact object mergers.\\

Finally, as a free by-product, we have explicitly shown the massive impact that Bayesian priors have in the parameter inference of high-mass mergers that display a very limited number of inspiral cycles, which leaves us with very little information about the binary parameters. We have shown that for face-on cases (for which only one mode is clearly visible), Bayesian priors can drive important biases in parameter estimation. These are, in contrast, greatly corrected when more than one mode is visible in band (\ie for unequal masses and large orbital inclinations), even if only the ringdown emission is observable. On the one hand, this shows once again the importance of higher-order modes in parameter inference of high-mass mergers~\cite{Graff:Missing_Link}. On the other hand, in our view, this also shows the importance trying a large suite of Bayesian priors to assess the robustness of the results for this type of sources~\cite{GW190521_Nitz,proca_21g,proca_obs}.

To close our work, we note that our study should be considered as a proof-of-concept that qualitatively assesses the detectability and impact of dense environments, as it is of course complicated to map the simple environment we chose to more realistic ones as e.g. Active Galatic Nuclei~\cite{Sberna:AGN_LISA_2022,Toubiana:AGN_LISA_2021}. Even in this situation, we believe our work constitutes a notable advance, as similar studies to date have considered modifications in the inspiral waveform obtained through post-Newtonian theory~\cite{Vitor:Env_2020,Vitor:Env_2022}. While this is extremely useful for the case of long-lived signals, this misses the important impact of the environment in the merger-ringdown signal we show thanks the usage of full numerical relativity. We hope that our work will contribute to further triggering the systematic performance of further numerical relativity simulations of black-hole mergers in realistic environments, as \eg those in~\cite{Fedrow2017}.

\section{Acknowledgements}
We thank Shubhanshu Tiwari for useful comments on our manuscript.
JCB is funded by a fellowship from ``la Caixa'' Foundation (ID100010434) and from the European Union’s Horizon 2020 research and innovation programme under the
Marie Skłodowska-Curie grant agreement No 847648. JCB is also supported by the research grant PID2020-118635GB-I00 from the Spain-Ministerio de Ciencia e Innovaci\'{o}n. The fellowship code is LCF/BQ/PI20/11760016. This work was supported in part by NSF awards 2207780, 2114581, and 2114582 to PL. We acknowledge the use of IUCAA LDG cluster Sarathi for the computational/numerical work. The authors acknowledge computational resources provided by the CIT cluster of the LIGO Laboratory and supported by National Science Foundation Grants PHY-0757058 and PHY0823459; and the support of the NSF CIT cluster for the provision of computational resources for our parameter inference runs. This material is based upon work supported by NSF's LIGO Laboratory which is a major facility fully funded by the National Science Foundation. This manuscript has LIGO DCC number P2300236. \\

\appendix
\section{Investigating waveform systematics}

\begin{figure*}[htbp]
    \centering
    \includegraphics[width=\textwidth]{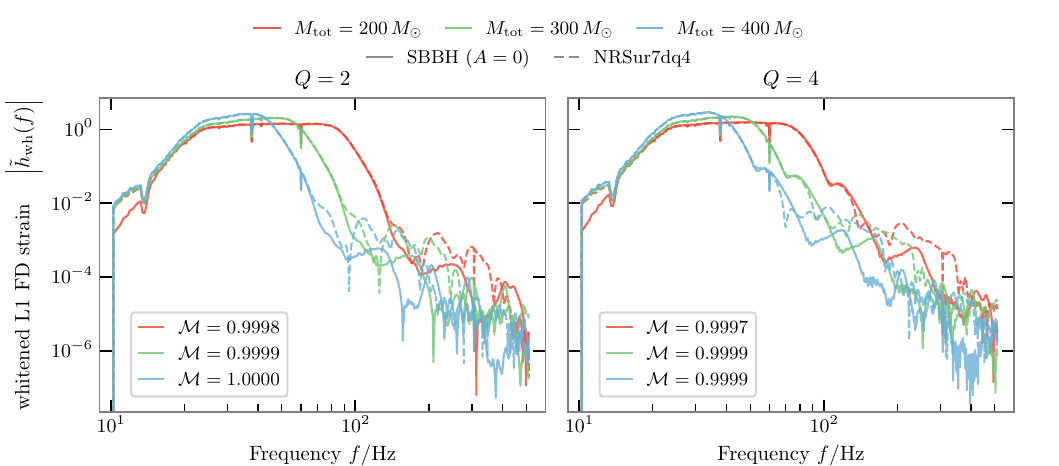}
    \caption{{\bf Match between our BBH waveforms and \nrsur} The solid and dashed curves respectively represent the amplitude of waveforms computed through our numerical relavitity simulations (solid) and by \nrsur{} (dashed), expressed in the Fourier domain. These are shown for several values of the total mass and whitened by the power-spectral density considered in the main text. The legend displays the match (in Eq.~\eqref{eq:matchSNR}) of each pair of waveforms. The left and right panels respectively correspond to mass ratios $Q = m_1 / m_2 = 2$ and $Q = 4$.}
    \label{fig:FO_match}
\end{figure*}

In this section, we check that our numerically simulated signals do not introduce further systematic differences with the waveform model \nrsur{} beyond those due to the presence of the scalar field. Fig.~\ref{fig:FO_match}, shows the amplitude of whitened frequency domain strain, \ie $\abs{\tilde h_{\rm wh}(f) }= \abs{\tilde h(f) / \sqrt{S(\abs{f})}}$, of our BBH waveforms ($A = 0$) and the corresponding BBH waveform obtained from \nrsur, together with their match (Eq.~\eqref{eq:matchSNR}). For each mass-ratio, we show the value of the match between the waveforms. We find a \textit{minimal match} of ${\cal{M}} = 0.9997$.

\section{Supplementary plots}

\begin{figure*}[htbp]
    \centering
    \includegraphics[width=0.95\textwidth]{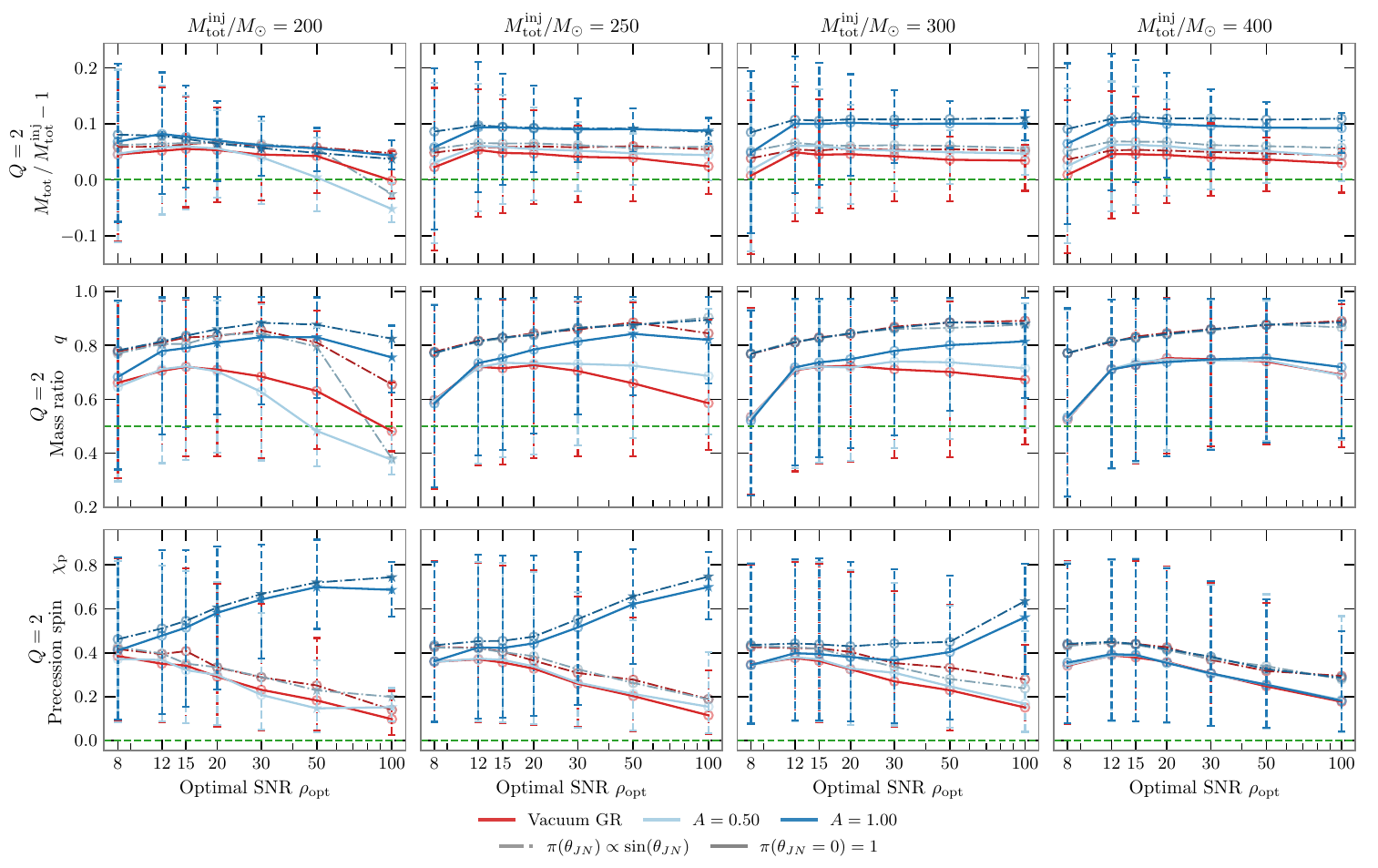}
    \includegraphics[width=0.95\textwidth]{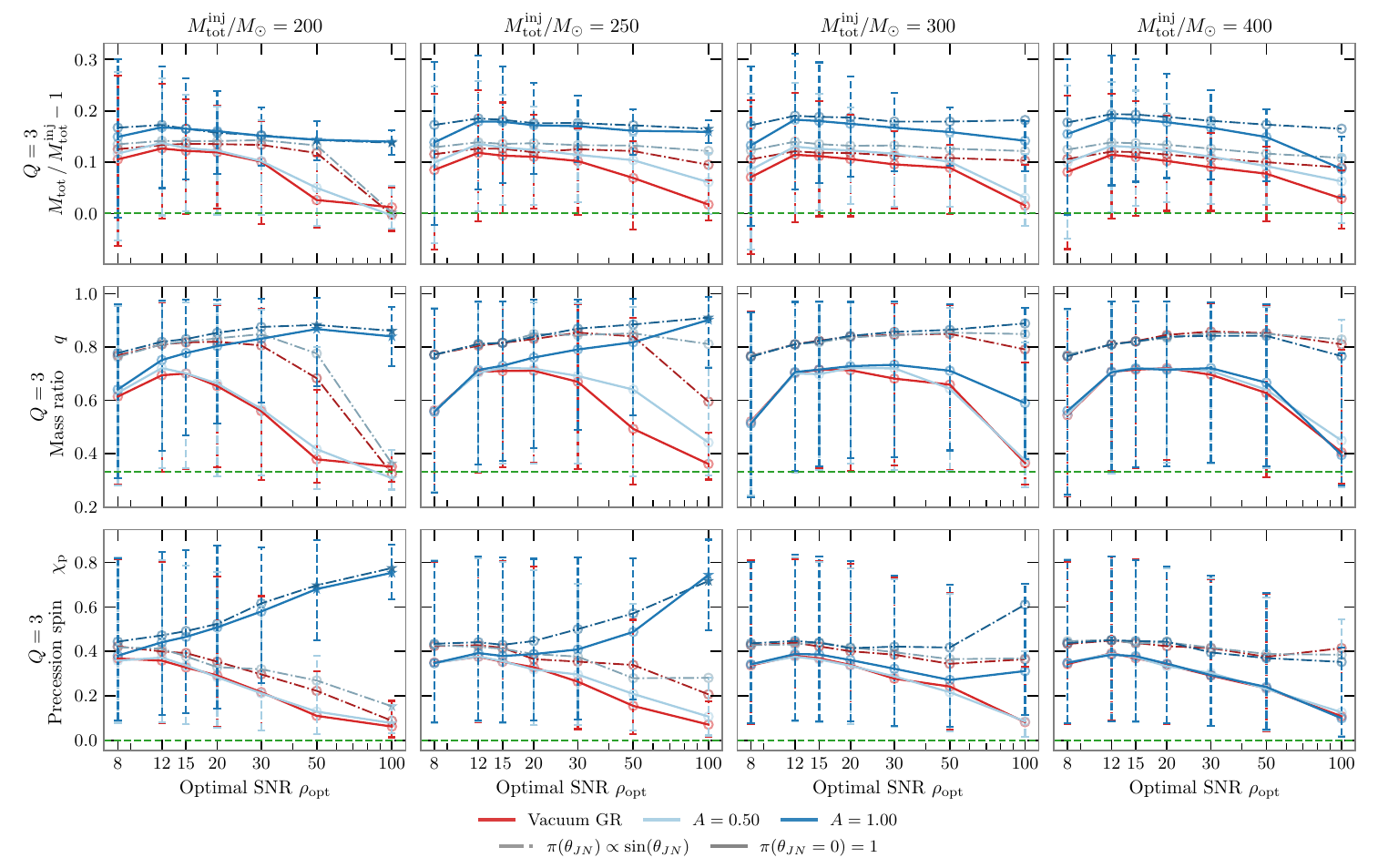}
    \caption{\textbf{Parameter biases for face-on binaries due to the omission of environmental effects, $\boldsymbol{Q = 2,3}$.} Same as Fig.~\ref{fig:faceon_bias} but for SBBHs with mass ratios $Q=2$ and $Q=3$.} 
    \label{fig:faceon_bias_q23}
\end{figure*}

For completeness, Fig.~\ref{fig:faceon_bias_q23} shows the same as Fig.~\ref{fig:faceon_bias} in the main text, but for cases with mass ratios $Q=2$ and $Q=3$. Both of them shows similar qualitative features as in the case with $Q = 4$.

\bibliography{bibliography,NumRel,matter_env,LIGO_papers}

\end{document}